\documentclass[a4paper,twocolumn,showkeys,floatfix,aps,pre,reprint,groupedaddress]{revtex4-1}
\usepackage{color}
\usepackage{amsmath,amssymb}
\usepackage{graphics,graphicx}
\usepackage{dcolumn,bm}
\usepackage{psfrag}
\usepackage{enumerate}

\begin{document}

\title{Long route to consensus: Two stage coarsening in a binary choice voting model}

\author{Sudip Mukherjee${}^{1,2}$}
\email{sudip.mukherjee@saha.ac.in}
\author{Soumyajyoti Biswas${}^{3}$}
\email{soumyajyoti.b@srmap.edu.in}
\author{Parongama Sen${}^{4}$}
\email{psphy@calunic.ac.in}
\affiliation{
${}^1$Department of Physics, Barasat Government College, Barasat, Kolkata 700124, India \\
${}^2$Saha Institute of Nuclear Physics, 1/AF Bidhannagar, Kolkata 700064, India \\
${}^3$Department of Physics, SRM University - AP, Andhra Pradesh - 522502, India \\
${}^4$Department of Physics, University of Calcutta, 92 Acharya Prafulla Chandra Road, Kolkata 700009, India.
}

\date{\today}

\begin{abstract}
  Formation of consensus, in binary yes/no type of voting, is a well defined process. However, even in presence of 
clear incentives, the dynamics involved can be incredibly complex. Specifically, formations of large groups of 
similarly opinionated individuals could create a condition of `support-bubbles' or spontaneous polarization that renders consensus virtually 
unattainable (e.g., the question of the UK exiting the EU). There have been earlier attempts in capturing the dynamics of
consensus formation in societies through simple $Z_2$-symmetric models hoping to capture the essential dynamics of average behavior
of a large number of individuals in a statistical sense. However, in absence of external noise, they tend to reach 
a frozen state with fragmented and polarized states i.e., two or more groups of similarly opinionated groups with frozen dynamics.
  Here we show in  a kinetic exchange opinion model (KEM) considered on $L \times L$ square lattices, that
while such frozen states could be avoided, an exponentially slow 
approach to   consensus is  manifested. 
Specifically, the system could either reach consensus  in a time that scales as $L^2$  
or a long lived metastable state (termed a domain-wall  state)
for which  formation of consensus takes a time   scaling as $L^{3.6}$. 
The latter behavior is comparable to some voter-like models 
with intermediate  states studied previously.
The late-time anomaly in 
the time scale is reflected in the persistence  probability of the model. Finally, the interval of zero-crossing of the 
average opinion i.e., the time interval over which the average opinion does not change sign is   shown to follow  a scale free distribution, which is compared with that
seen in the opinion surveys regarding Brexit and associated issues in the last 40 years. 
The issue of minority spreading is also addressed by calculating the exit probability.
\end{abstract}


\maketitle


\section{Introduction}
The dynamical  evolution of opinion formation in a society is a complex process involving myriads of socio-economic,
not to mention psychological, issues. Even the quantification of opinions by numbers, therefore, could be
difficult to define, let alone their evolution \cite{stauffer,sen_chak,soc_rmp,galam_book}. However, when the
choice is binary, e.g., in a yes/no referendum, a two party voting etc., the said quantification is straight-forward. It is
often done with $\pm 1$ values, i.e., like spins with Ising symmetry, for example in the voter model \cite{cliff,ligg1,ligg2}. Additionally, it could be crucial to have
neutral opinions - with opinion value zero - reflecting the population that does not belong to either of these groups \cite{bcs_us}.

In spite of the complexity involved, the
literature over the last three decades or so indicates, at least in the statistical physics community,  
that the essence of the opinion evolution in a society 
could be captured by models that assume that  the change in the individual opinion  is activated by  interaction between two or few agents \cite{sen_chak}.
The diversity can  come through  various factors. The choice of the partner for interaction can be crucial. Most 
of the models consider only   nearest neighbors interactions. But there can be additional restrictions, 
e.g., for the bounded-confidence models, where the interaction is conditional upon a limited range of the absolute difference of opinions 
of the two individuals \cite{def1,heg}.
One can consider  kinetic exchange models (KEM), where   the 
relative weights of one's own opinion
and that of their peers  \cite{tos1,lccc,psen2010,soumya2011,psen2012,soumya2014,bcs,tos2} are both decisive factors in the evolution of the opinions.
Unlike the one-to-one interaction of the KEM, to take peer-pressure into account, in some models, a group of individuals 
having identical  opinion can together affect one individual opinion, as in \cite{szj,soham}.
The topology and/or the hierarchy of the neighborhood of the agents (social-networking)  also play a key role by limiting the interaction range.
Of course, many other factors can be incorporated  which add to the intricacy of the models 
but are beyond the  scope of discussion
in the context of  the present paper (see for a recent review \cite{sen_chak}). 

Each of these variations has its own value in representing certain global or average feature of the opinion formation
in a society of a large number of interacting agents. However, in terms of formation of an overall consensus \cite{cons}, these models, especially
in dimensions two or higher, are often inadequate. 
Many of these  models reach a frozen or a fragmented state in the long time.  
 The dynamics after that either cease to continue  or evolve in a way that no further
movement towards consensus is possible. Examples range from the two state  models  like  
the Ising model (which can be regarded as a simple opinion dynamics model)
in two dimensions  \cite{b_redner} and both the Ising and voter models on well-studied 
complex networks \cite{svenson,hagg,boyer,castellano,biswas,baek,khaleque,krap-comp,castell2003,sood-red2005,such-2005-1,castellano} 
to the bounded confidence models with  multistate or continuous opinions  
(see e.g., \cite{def1,heg,naim,vaz}).

Here we show  that in a kinetic exchange model of opinion formation on square ($L \times L$) lattices,  which allows neutral opinions in an 
otherwise binary choice situation, a consensus is always reached, but it is reached in a two-stage process. 
In the first stage, starting from the random initial conditions, we note the usual Ising-like coarsening dynamics, i.e., the order parameter scales as $t^{1/2}$. 
Beyond a time scale $\tau_1 \propto L^2$,  the order parameter shows a different behavior with time as, with a finite probability,  the system reaches a  
 so called domain-wall state.  In the two dimensional Ising coarsening,
all initially random states do not reach consensus but some get stuck in a locally stable 
state characterized by rectangular domains of $\pm 1$ state. Here, 
states which consist of  domains with nearly straight edges are  reached somewhat reminiscent of the frozen states in the Ising model. 
Although these states are long-lived, the domain walls are eventually unstable.  
The system always evolves to a consensus state, but in a much longer time-scale,
 $ \tau_2 \sim L^{3.6}$. 
It is in this long metastable state
that the two competing, relatively similarly sized domains of opinion values, continue to evolve, showing a scale-free 
scaling of the size of the intervals within which the society as a whole switches the sign of majority (zero-crossing probability of the average opinion).

Since the KEM in two dimensions
 belongs to the Ising universality class as far as equilibrium properties are 
concerned \cite{nuno1,nuno2,sudip,lima}, 
 one of the  main concern is whether/how the dynamics are different from the kinetic Ising model.  
Having said that, it is interesting to note that  the domain wall states that characterize the latter part of the dynamics of the three state KEM model are   
comparable to similar such states observed in  language dynamics models \cite{migu,cast}  and voter-like models   \cite{luca,volo,vazquez} where 
intermediate (language or opinion) state(s) are present. Indeed, in some of these cases, the   consensus time  scales with the system size
with an exponent  
 value close to 3.6 that we get in the present work. 
Thus the slow coarsening time scale seems  generic in the cases with
intermediate choice(s) although the  microscopic dynamical rules  may be quite different.  
 We will elaborate this point further in sec. IV, while comparing the present KEM model with other models.


To continue the comparison with the Ising dynamics, we have computed a few other dynamical features 
e.g., persistence probability and  exit probability  discussed in the next two sections;  both  are very well studied quantities for spin models.

\begin{figure}[tbh]
\centering
\includegraphics[width=8cm, keepaspectratio]{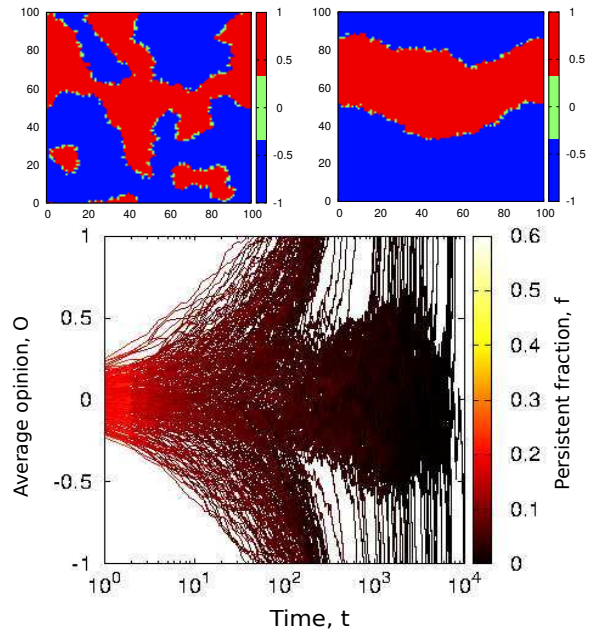} 
\caption{ Evolution of the average opinion values with time.
Upper panel: The two snap shots, taken at $t=10$ (left) and $t=1000$
(right), show the domain walls at those times. Clearly, for the configurations where consensus was not reached at an early stage, 
a domain boundary is formed. The boundary is populated with neutral agents, as follows from the model dynamics, and therefore
the domains are never frozen. Lower panel: Average opinion as a function of time shows the signature of the two time scales. The color/shade on the lines represent the fraction of persistent sites (those never flipped) 
up to that point.}
\label{schematic}
\end{figure} 


Finally, for comparison with a similar situation in society, we turn to the opinion surveys for the last 40 years in the UK
regarding the issue of exiting the European Union (which recently has come to be referred to as Brexit). We note that this is 
indeed a situation where  similarly sized domains continue to evolve. We compare the interval distributions of the zero-crossing
time of the average opinions (intervals to switch between ``leave" to ``remain" majority) with that computed in  our model.

\section{Model and salient  dynamical features}

The kinetic exchange model for opinion formation \cite{bcs} is considered on a square lattices where each lattice site is 
occupied by an agent.  
The  
opinion of any agent can have three values: $\pm 1$ or zero. 
For the  agent at  site $i$ 
(called the $i$th agent),
 interacting with the $j$th agent (chosen randomly from one of the four nearest neighbors of the $i$th agent),
the opinion value $o_i$  changes according to
\begin{equation}
  o_i(t+1)=o_i(t)+\mu_{ij}o_j.
\end{equation}
No sum over the index $j$ is implied, as the model is binary-exchange.  A non-linearity enters the model from the
imposed bounds in the opinion values for the extreme ends at $\pm 1$, i.e., $|o_i|\le 1$, signifying the limit to an extreme opinion.
For this work, $\mu_{ij}=1$, while originally \cite{bcs} it is a stochastic variable having both positive and negative values (the code used here is freely available \cite{code}).
The fact that $\mu_{ij}=1$ always, indicates a conducive environment for consensus formation, in the sense that the difference of opinion between two interacting agents either remain unchanged or decreases following the interaction.
Clearly, if the initial condition of the model is restricted to the agents having opinion values $\pm 1$ and $0$, for all subsequent 
times, the opinion values are confined within these three values.  For a binary choice case,   
the two given choices, say `leave' and `remain' (e.g., in Brexit) are represented by opinion states $\pm 1$ and the agents who   remain neutral/indifferent are assigned opinion ``0".  
This steady state  properties of this model  have  been studied in mean field and other topologies to some extent (see e.g., \cite{nuno1,nuno2,sudip,lima}).
A trivial fixed point of the model is when all $o_i=0$,
but this is unstable with respect to any single agent  with  a nonzero opinion being present. Given that our initial conditions never start with all
neutral  agents, this state is never reached.  

 Although the opinions can take three possible values  1, 0, -1 in the kinetic exchange model,  only the `all 1' /`all -1' states are the  fixed points of the dynamics 
whenever the initial state has at least one agent with nonzero opinion value. 
Our main results concern the route to the consensus states and in addition we have estimated  the persistence probability and exit probability which are 
defined in the following.

 The persistence probability \cite{satya-rev} $P(t)$ 
is defined in general as the  probability that 
a field has not changed sign till time $t$. 
In opinion dynamics (spin) models, it is the fraction of  opinions (spins) 
that  remain unchanged till $t$. In the zero-temperature Ising Glauber model in two dimensions, it is known to behave as $P(t) = t^{-\theta}$ with $\theta$  very close to 0.20 \cite{blanch}. 

In general we study the dynamics starting with an uniform distribution of opinions with average opinion equal to zero (the most noisy state). 
One can also choose a biased initial configuration  
by taking 1/3 of the opinions equal to zero, $\frac{1}{3} + \frac{\Delta}{2}$ fraction equal to 1 and the rest equal to -1. Then the exit probability $E(\Delta)$ measures the 
probability that the system reaches a `all 1' state eventually. It is interesting to know   whether it is possible  
to  reach a  consensus state with all opinions equal to +1 
starting from a state with majority of  agents having opinion = -1.  Mathematically, this requires $E(\Delta) > 0$ for  
$\Delta < 0$.  
This kind of reversal of initial opinion is similar to the phenomenon of   `minority spreading' \cite{galam-ms}.

Since $\mu_{ij}=1$ always, after a transient state, the neutral opinions are exclusively confined at the domain boundaries of 
opposite signs. This is because  neutral agents necessarily interacting with agents having extreme opinion values, are unstable in the sense that it always changes opinion values when interacting with an agent with $\pm 1$ opinion value. 
It is worthwhile to note at this point that as a consequence of the above observation, in one dimension, the dynamics of this model are 
 equivalent to that of the zero temperature coarsening of the Ising model using Glauber dynamics. The only difference is the presence of domains of neutral opinions  in the kinetic exchange model.  
However, such domains are limited in their life-spans that are proportional to
their initial lengths which eventually makes the two models behave in the same manner.
As long as the lower dimensions are
concerned, the first occurrence of a non-trivial difference in the dynamics of the model is for dimension $d=2$. The reason for this is straight-forward. 
Given the dynamical evolution rules of
the model, the neutral agents confined between the domains of the extreme 
opinions keep the dynamics alive, as long as domain boundaries exist. This means that 
the system is  dynamically active   
 until a consensus is reached. 
However, the time needed for reaching a consensus as a result of this `active-boundaries'
is much higher than the usual coarsening time. 

%
%

\begin{figure*}[tbh]
\centering
\includegraphics[width=8cm, keepaspectratio]{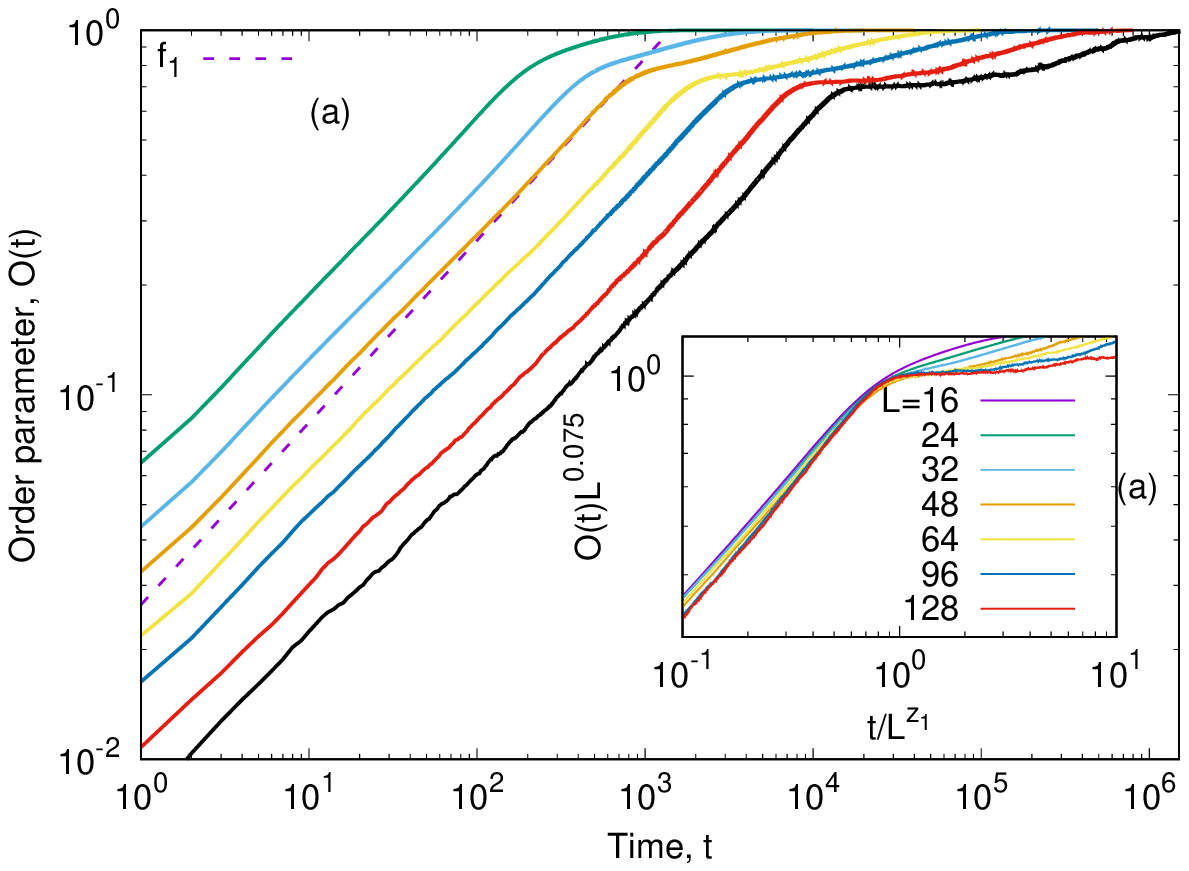} 
\includegraphics[width=8cm, keepaspectratio]{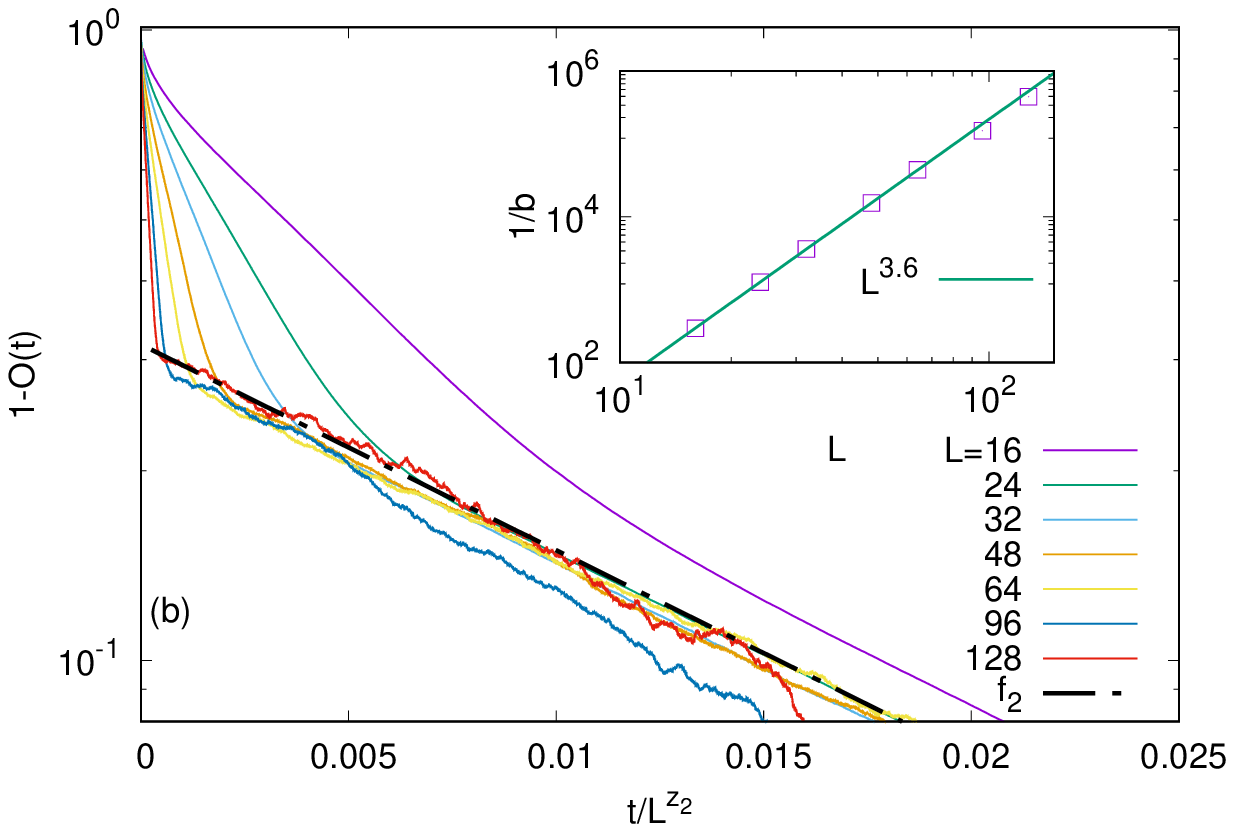} 
\caption{The growth of the order parameter with time. The change in slope in the log-log plot shows the existence of the 
two time scales for consensus. The first time scale collapses when the X-axis is scaled by $L^2$ (a), where
the second time scale (final consensus time) remains uncollapsed. The growth of the order parameter follows $f_i\sim t^{0.50\pm0.01}$.
 Again, the second time scale collapses when the X-axis
is scaled by $L^{3.60\pm0.01}$ (b), where the first scale remain uncollapsed. The form of $f_2$ is given in Eq. (\ref{fitting})}.
\label{two_scales}
\end{figure*}

\section{Numerical results}

A qualitative idea of the dynamics can be gained  by looking at the 
 opinion configurations of the KEM at various stages of the dynamics (see Fig. \ref{schematic}).
The snapshots  clearly indicate the presence of two time scales. 
At the initial times ($t<\tau_1$), the spin domain configurations are random. At a later stage ($\tau_1<t<\tau_2$), if the system  has not already entered a fully consensus
state, it is in a state with two (or more, in principle) large domains (taking the periodic boundary conditions into account).

\subsection{System size scaling of the two consensus time scales}

We first study the behavior of the global order parameter given by 
$O(t) = |\sum\limits_i o_i(t)|/N$. 
The simulations are done on square lattices, with the sizes ($N=L\times L$) indicated in the figures and following periodic boundary conditions. Unless otherwise mentioned, the initial conditions are completely random i.e., equal probability for the three opinion values $\pm 1$ and $0$. The realization averages are between 10000 to 100, depending on the system sizes.
In Fig. \ref{two_scales} we show  how 
the order parameter evolves in time for various system sizes. It is clear from 
Fig. \ref{two_scales}(a)  that the evolution shows the effect of the two time scales
mentioned earlier. For the initial part of the dynamics, the order parameter increases in a power-law manner with time, with 
the exponent value about $0.50\pm0.01$, which
is  the same as in Ising model coarsening. The growth, however, drastically slows down when the second time scale manifests
itself. The subsequent growth is much slower, which continues until complete consensus is reached. The inset of Fig. \ref{two_scales}(a)  confirms 
that $\tau_1 \propto L^{z_1}$ where $z_1 =2$ is a dynamic  exponent of growth. 
This value of $z_1$  indicates a  dynamical evolution which  is
curvature driven and valid for the 
zero temperature coarsening of  Ising model in all dimensions.  

For $t > \tau_1$, the order parameter  
can be  fitted to a functional form 
\begin{equation}
O(t) \sim 1-a \exp(-bt),
\label{fitting}
\end{equation} 
indicating  that $1/b$ represents another time scale which we identify as $\tau_2$. 
$1/b$ is   shown in the inset of Fig. \ref{two_scales}(b) and we find
\begin{equation}
1/b \sim \tau_2 \propto L^{z_2}  
\label{z2}
\end{equation}
where $z_2 = 3.60 \pm 0.01$. 
In Fig. \ref{two_scales}(b)
we obtain a partial collapse of the data for the late time regime using 
using Eqs (\ref{fitting}) and (\ref{z2}).


\begin{figure}[tbh]
\centering
\includegraphics[width=8cm, keepaspectratio]{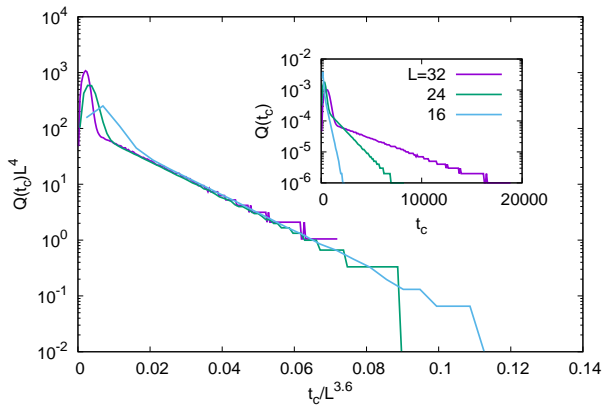} 
\caption{Consensus time distribution and its finite size scaling. The inset shows the distribution function $Q(t_c)$ of the consensus time ($t_c$) for systems of sizes $L=16$, $24$ and $32$. 
The shorter time scale ($\tau_1$) is manifested through a peak in the
distribution (configurations reaching consensus without forming two domain states), while the configurations with two 
domains contribute to the exponential tail of the distribution. The data collapse of the exponential 
part in the main figure recovers the scaling $\tau_2\sim L^{3.6}$.}
\label{absorbing_time}
\end{figure}
     
\begin{figure}[tbh]
\centering
\includegraphics[width=9cm, keepaspectratio]{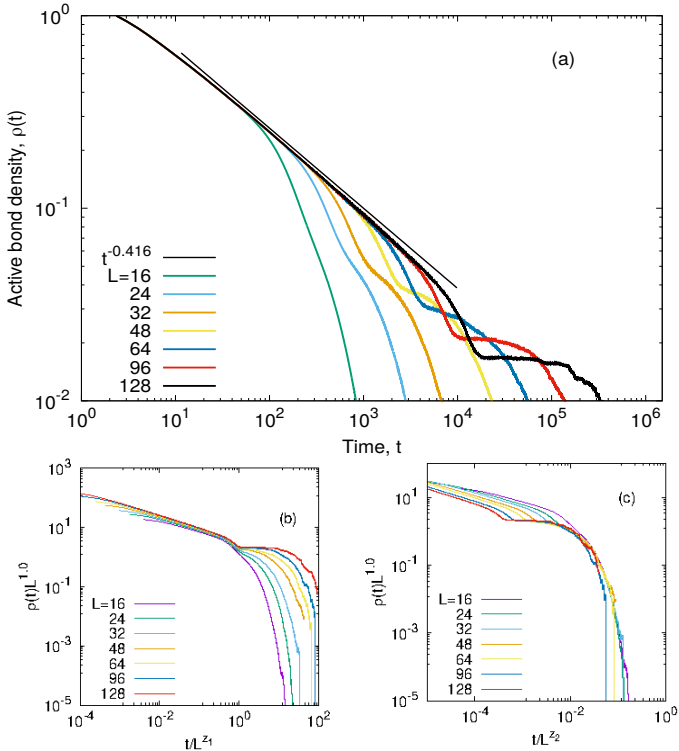} 
\caption{Density of unlike neighbors (active bonds) and its finite size scaling. (a) The density of active bonds (pairs of neighboring unlike opinion values) are plotted for various system sizes with time. After an initial
power law decay, it reaches a plateau, which corresponds to the domain wall states. It finally drops to zero exponentially. (b) The time at which
the plateau is reached should correspond to the time at which domain walls appear, hence a rescaling of the time axis by $\tau_1\sim L^{z_1}$ shows 
a data collapse. On the other hand, the value of the bond density at the plateau should scale as $1/L$, given the domain wall nature of the configuration.
Therefore, rescaling $\rho(t)$ by $1/L$ shows data collapse. (c) Finally, the time scale appearing in the exponential part should correspond to $\tau_2\sim L^{z_2}$, hence a rescaling of the time axis by $\tau_2$ gives a data collapse.} 
\label{active_bonds}
\end{figure}

We also obtain the distribution $Q(t_c)$ of the 
 consensus times $t_c$ as shown in  the inset of Fig. \ref{absorbing_time}. 
 The peak in the distribution function corresponds to the shorter consensus time or domain-wall formation time
scale $\tau_1$, where the configurations either reach consensus avoiding the domain wall formation or reach the domain-wall states. However, the tail of the distribution, which 
represents the time scales for the domain walls to merge, i.e., formation of the overall consensus, gives the second, much larger, time scale $\tau_2$.
One can obtain a collapse of the data for the tail by 
 scaling $t$ by $L^{3.6}$. It is also observed that the scaling function has
an exponential behavior. 

\begin{figure}[tbh]
\centering
\includegraphics[width=8cm, keepaspectratio]{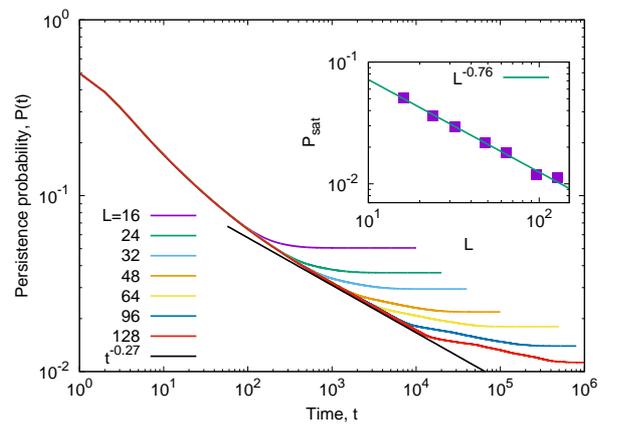} 
\caption{The persistence probabilities are shown for the two dimensional version of the model. The exponent value for the decay, $0.270\pm0.001$, 
is distinct from what is obtained in the two dimensional Ising model \cite{blanch}. Inset shows the variation of the saturation values of $P(t)$ against the system sizes.}
\label{2dbcs_perst}
\end{figure}

\begin{figure}[tbh]
\centering
\includegraphics[width=9cm, keepaspectratio]{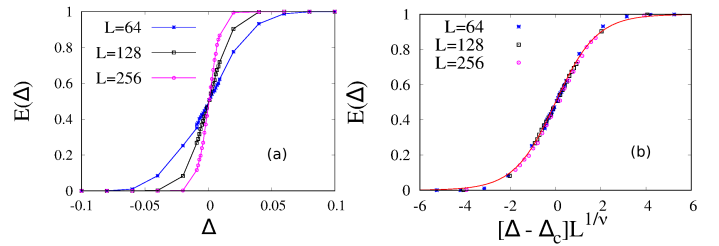} 
\caption{The exit probabilities for the model are shown for different system sizes (a) and then a finite size scaling was done (b).
The finite size scaling exponent ($\nu=1.05\pm 0.01$) is distinct from what is seen in the case of Ising model ($1.25$) \cite{parna}.
The collapsed curve fits with a function of the form $f(x)=[1+\tanh(cx)]/2$ with $c=0.547\pm 0.015$.}
\label{bcs_exit_prob}
\end{figure}

Finally, for the analysis of the two time scales, we look at the active bond density $\rho(t)$, which is defined as the density of neighboring pairs of agents having different opinion values. This quantity is widely studied in the voter-like models, and  shows a power law decay when intermediate 
states are allowed. We note a power law decay here  in the initial stage with an exponent value $0.416\pm0.001$. 
The decay then slows down and shows a plateau region which is 
attributed  to the formation of the domain
wall states   
 (see Fig. \ref{active_bonds}), $\rho(t)$   eventually drops to zero, when a consensus is reached. The time scale for reaching the plateau and then the eventual exponential decay should correspond to the two time scales $\tau_1$ and $\tau_2$ obtained before.  Data collapses in the initial and later stages,  obtained using the scaling variables $t/\tau_1$ and $t/\tau_2$ respectively and shown in Fig. \ref{active_bonds}, confirm  the presence of the two time scales. The collapses improve for larger system sizes.

\subsection{Persistence and exit probabilities}
The persistence of the opinion value of a given agent is an interesting measure of the stability of dynamics in a model. In zero temperature
quench of Ising model, this is well studied \cite{satya-rev}. The persistence probability $P(t)$ at any time is the fraction of agents who did not change
their opinion values at all up to that time. Fig. \ref{2dbcs_perst} shows the decay of the persistence probability for the KEM. Although it 
does not  show a clear  power law behavior for small system sizes, for the largest size, a power law behavior can be detected over a considerable time range.   From this regime,  one  obtains 
\begin{equation}
 P(t)\sim t^{-\theta}, 
\end{equation}
with $\theta \approx 0.270\pm0.001$,  
clearly  distinct from that found in the two dimensional Ising model \cite{blanch}. 
Furthermore, due to the fact that the dynamics may not be frozen after time $\tau_1\sim L^2$, the persistence probability does not become time independent beyond $L^2$ (before reaching a saturation value) as in the 
case of Ising model. However, the subsequent decay of the persistence probability is, like the growth of the order parameter discussed before, very slow.  
Hence, the effect of the two time scales described above is also visible in this measure. 
We also note that the saturation value of $P(t)$ varies with $L$ as $L^{-0.76\pm0.01}$ shown in the inset of Fig. \ref{2dbcs_perst}.

The exit probability is another important measure that quantifies the probability  to reach a particular 
consensus state with an initial bias. In particular, if an initial bias  is ineffective  (e.g., an initial  configuration 
 with more $-1$ ultimately ends up in an all $+1$ state) it can be called a case of minority spreading.  
 This is again a very well studied quantity in Ising model and other opinion dynamics models \cite{sznajd_exit,sznajd2,lamb,timp,parna,parna1,
parna2,suman}. 
The general scaling form of the exit probability $E(\Delta,L)$ reads
\begin{equation}
E(\Delta, L)=\mathcal{F}\left((\Delta-\Delta_c)L^{1/\nu}\right)
\end{equation}
where $\Delta$ is the bias in the initial opinion fractions (initial densities are $\frac{1}{3}$, $\frac{1}{3}+\frac{\Delta}{2}$ and $\frac{1}{3}-\frac{\Delta}{2}$ for opinion
values 0, $\pm 1$ respectively; $-2/3 \leq \Delta \leq 2/3$), with $\Delta_c=0$ and $\nu$ being interpreted as a  correlation length exponent
(to be distinguished from the critical correlation length). 
The 
exit probabilities and their finite size scaling are shown in Fig. \ref{bcs_exit_prob}. The exponent value of $\nu$ for the collapse is
$1.05\pm 0.01$ which is again distinct from what is
seen in the Ising model in two dimensions ($ \approx 1.26$) \cite{pratik2017}. 
The scaling function can be fit to the form $f(x) = [1+ \tanh (cx)]/2$ with $c = 0.547\pm 0.015$.

\begin{figure}[tbh]
\center
\includegraphics[width=9cm, keepaspectratio]{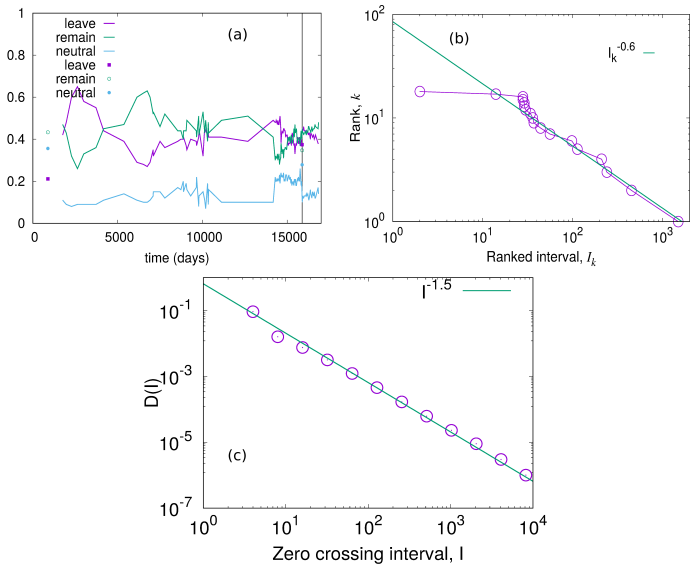} 
\caption{COmparisons between Brexit and model data.  (a) The results of various opinion surveys and referendum on the question of the UK leaving the EU from the date of its joining (1 January, 1973 to the then European Communities) as the origin ($t=0$) are shown. The
vertical line denotes the time of the last referendum (23 June, 2016). (b) The rank-plot of the interval of the zero-crossing of the net opinion value (difference
between remain and leave fractions) is shown. The tail of the rank plot shows an exponent close to $-0.60\pm0.02$. (c) The distribution $D(I)$ of intervals of zero crossing for KEM.}
\label{brexit_data}
\end{figure}

\section{Comparison with other dynamical models}

The present model, although a three state model, has only two stable
fixed points, all +1/ all -1 and hence can be compared with other $Z_2$ models
studied extensively in the literature. We have in mind precisely the 
Ising model or   voter like models which are special cases of the generalized voter model \cite{oliv}.   
We have already mentioned some  differences compared to the Ising model,
the most important of which is  the absence of frozen states at extremely large times. This is attributed to the fact that the zero states are primarily 
 found at the domain boundaries which on interacting with any nonzero opinion is
bound to change. So the boundaries of a zero domain naturally recede resulting into a narrowing of the zero band before it vanishes. The vanishing of a zero 
domain brings into contact domains of $\pm1 $ opinion thereby again producing some zeroes at their boundary. However, as in any finite system, a random fluctuation will drive the system to either of the all 1 or all -1 state. While the consensus state is reached for all configurations, this kind of slow dynamics to reach the consensus state is seen in about  30$\%$ of all the random initial configurations used.
Such metastability is possible 
only in dimensions greater than one, hence in one dimension,    dynamics of both Ising model and KEM  coincide. It is already 
known that the voter model dynamics are  identical to that of Ising in one dimension, so that all three models show the same dynamical behavior
in one dimension.  



In the KEM, as discussed before, there are two time 
scales in the coarsening process, the second being much larger than the first.
It is true that in the two dimensional Ising model also, one gets a
two time scale scenario due to the slow relaxation of the so called diagonal states \cite{b_redner} where the scaling of the second time scale is quite similar to
what we get. However, such cases were much more rare (nearly 5$\%$ of the consensus states)  to 
affect the bulk dynamical behavior  compared to   KEM where the 
slow dynamics occur in about $30\%$ cases.
 In the Ising model, the dynamics are always curvature driven. In the KEM,  the initial evolution is perhaps the same  indicated by the value of $z_1 =2$. 
On the other hand, in the voter model, the dynamics are interfacial 
noise driven so that it shows a different dynamical behavior in two dimensions. In the KEM, the dynamics are like the 
voter model only for the zero state which simply copies the state of the interacting agent and in that sense partially interfacial
noise driven. Since at a later stage, it is basically the zeros which undergo a change, the dynamics become more and more 
 interfacial noise driven. 
In the generalized voter model, the interplay of the   interfacial 
noise and curvature driven dynamics was studied recently \cite{parna2} which showed   crossover from a faster to a 
slower dynamics due to the  presence of   metastable states with   similar nearly straight-edged domains. It was the interfacial noise
which was found to lead the system to consensus in the later stages and one can argue that the same is happening in KEM. 
In the generalized voter model also,  two time scales were detected, 
however, such a  large value of $z_2$ was   not indicated in that case. 
On the other hand, the exponent value $z_2 \sim 3.6$ obtained  
in the KEM  is
quite close to that found in a  two state majority rule model \cite{chen} where also one gets long lived metastable states. 
In this model,  the signature of the  slow relaxation was captured by the behaviour of the 
consensus time distribution from which   $z_2  \approx 3.4$ was obtained by extrapolation   in two dimensions. A scaling argument, however,  indicated that the   value should be equal to 3.

Slow coarsening, where the time scale varies with $L$ with an exponent close to 3.6, has  also  been  observed in some voter-like models, where one or more intermediate states exist between the two 
extreme values of the opinions. These include the model for bilingual
dynamics, three or more state  voter models etc. \cite{migu,cast,luca,volo,vazquez}.   
In these studies, the intermediate states 
introduced an effective surface tension in the domain walls, driving the system to a curvature driven 
state. While the concentration of unlike pairs decays approximately as $t^{-0.5}$ like in our case (we obtained a value $\sim 0.416$), 
the scaling $\tau_2 \propto L^{3.6}$ is found by not only considering the 
consensus time distribution but explicitly  from 
  the exponentially slow approach to the consensus 
state manifested by both $O(t)$ and  $\rho(t)$ in the later stage only in the present work. 
Except for some special cases for a four state model in \cite{vazquez},  where  
an  exponential slow approach   of the order parameter 
in the later stages was noted and was remarked to be unique, such a behaviour has not been reported earlier to the best
of our knowledge. Thus we argue that the mechanism in the later stages is 
interface noise driven in the KEM in contrast to that in the three state  language  and voter like models where the signature of a two stage process has  been reported only from the behaviour of the exponential tail of the consensus time distribution and not from the behaviour of the order parameter relaxation.  

It is  easily checked that the microscopic dynamics in the KEM  is 
quite different compared to e.g., the three state voter model. For example, 
in  the  three state voter model, 
the probability that a zero becomes 1  is equal  to 
 $f_1 + f_0/2$
where $f_k$ is the density of the opinion $k$ of the four neighbours while 
in the KEM, it is equal to $f_1$. 
In this context, it can be mentioned that for the two state models, a generalised formulation in terms of the
transition probabilities involving  two parameters has been made \cite{oliv}. The  Ising-Glauber  model, which uses  an
energy minimising dynamics and the voter model, in which the agents simply copy the opinion of a randomly 
chosen neighbour,  correspond to  two different  points in this two dimensional parameter 
space. These two have quite distinct dynamical behaviour in two or higher dimensions with respect to 
coarsening, persistence, freezing probability etc.  For three state models, one can, in principle, conceive of a similar 
generalisation with  a presumably  larger dimensional parameter space.
The KEM and the other voter like models,   quite different  in terms of their microscopic dynamics,  are  expected to correspond to reasonably well separated points/regions  in this parameter space. 
Even then, the fact that the KEM and the other three state models all show the slow coarsening behaviour with similar dynamical exponent values, strongly suggests that 
 this particular feature is  quite generic. 
Such is, however, not the case in two state models, where it is meaningful to 
consider variations in the parameters and see the effect, as was done in \cite{parna2}.

In none of the multistate models,  results for  the scaling of persistence probability or the
exit probabilities are available, the two other studies which highlight the present work.  
We note that the persistence probability in KEM decays with an exponent that has a larger value compared to that in the Ising model. This is presumably because in the initial stages many of the zeros definitely change their state thereby reducing $P(t)$ in a rapid manner. However, the dynamics in KEM lasts longer and in the later regime, the persistence probability still decays, albeit much slowly as very few changes in state occur for the first time. In comparison, in the Ising model, the system reaches a steady state much faster and $P(t)$  attains a constant value after a time scaling as $L^2$.  On the other hand, the persistence probability decays  
slower in comparison  to the voter model (numerically it is found to be $\exp[-const (\ln t)^2$] \cite{krap-comp}) where 
 $d=2$  is a  critical dimension for which  the dynamics are  known to be slow {\it {for all configurations}}. 
Consistently,   the saturation value of the persistence, 
which signifies the end of the dynamical evolution, comes at a much later stage in KEM. This in turn, affects the scaling form of the persistence probability.
In Ising model the saturation value  varies as  $L^{-\alpha}$, with   $\alpha=\theta z$ \cite{manoj_ray}, 
where $z=z_1=2$ is the unique  dynamical exponent for the Ising model 
and  $\theta =0.2$.
But in KEM, the saturation value scales as $L^{-0.76}$, which does not satisfy the scaling relation mentioned above for either $z_1$ or $z_2$, with 
$\theta \approx 0.27$.  

The exit probability in the voter model is known to vary linearly with the bias $\Delta$ in all dimensions. 
 This follows from the global conservation of the opinions over all ensemble, also true for the one 
dimensional Ising model.  In two dimensions, the Ising model and the Sznajd model  show a behavior similar to what we obtain for the 
KEM, however,  the values of the parameters in the scaled functions for the Ising model are quite different ($\nu \approx  1.26$ and $c \approx  1.1$ \cite{pratik2017}). 
Although in the thermodynamic limit, the exit probability assumes a step 
functional form, for finite sizes, there is  evidence  of minority spreading as indicated by the presence of the universal 
scaling function, a feature shared by the Ising model, though in the latter it happens to a lesser extent as the value of $c$ is larger.

\section{Comparisons with data}
There are examples of issues that eluded consensus in the society for decades and formed a spontaneous polarization like the domain wall state discussed here. One recent
example is the question of the United Kingdom leaving the European Union - commonly referred to as Brexit. This issue has certainly 
made the UK and EU as a whole, more polarized. From opinion surveys dated back to 1970s \cite{data}, there seems to be a fluctuating tendency in the
opinion values, while no side got overwhelming support. These are qualitative features we also find in the KEM considered here. However,
while in the model we can  quantify these features with various measures, with the data of opinion surveys on Brexit (or related questions), such
quantification is difficult.

Nevertheless, we can look at the intervals of the so called `zero-crossing' of the overall opinion values, i.e., the intervals in which the majority opinion shifted between ``leave" and ``remain". Fig. \ref{brexit_data} shows a rank plot of the
intervals, which suggests an approximate exponent value of $-0.6$ for the cumulative distribution, which in turn suggests $-1.6$ as the 
exponent value for the probability distribution function \cite{clauset} of the interval of switching of opinions. 

For comparison, we considered the time series of the global order parameter of KEM only for the configurations that did not reach consensus
in $\tau_1$ i.e., those configurations which reached a domain-wall state. The distribution $D(I)$ of the zero-crossing intervals for those configurations only, 
gives an exponent value $-1.5$ (see Fig. \ref{brexit_data}), which closely resembles the value suggested from data. However, as noted before, the data available are
rather sparse in measurement intervals, which can affect the distribution especially in the lower-cutoff region (not shown). Nevertheless, a manifestation of an
active exchange of opinions between two polarized groups is seen in the data that was described in the model.  

A possible argument for the origin of the $-1.5$ exponent value in the model could be the diffusing motion of the domain walls in  
effectively one dimension. The exponent is what is seen in the surviving probability for such a walk \cite{redner}.

\section{Discussions and conclusion}

Formation of consensus in a multi-agent society has been long viewed as a collective emergent phenomenon. 
In various models that considers the phenomenon as a critical behavior, in presence of sufficiently weak
noise, a majority opinion builds up \cite{sen_chak,galam_book,def1}. 
Here we have shown, however, that even in the absence 
of external noise, formation of opinions can be hindered due to spatial constraint in interactions. Specifically,
in the kinetic exchange model of opinion dynamics studied here, we find that for about 30\% of realizations, the 
configuration of opinions in a two-dimensional lattice go to a segregated or domain-wall state, where there are two
large groups of opposing opinion values. The groups are, however, not frozen in time.  The domain walls move,
due to interactions mediated by neutral agents, and eventually merge to form a consensus state. The time-scale 
to reach the consensus, however, is significantly higher than the time scale for reaching consensus for the non domain-wall states. 
Indeed, the two time-scales scale differently with system size as well.
The scaling of the persistence probability and the overall order parameter quantify these two time scales
in the model. This behavior is distinct from other coarsening dynamics, say in two dimensional Ising model, 
where the domain wall state is frozen in time in absence of external noise. 

There are real world scenarios where formation of consensus takes exceptionally long time. It is long seen that
in presence of external noise, bounded confidence or zealots, inertial agents etc., formation of consensus could be hindered. But even
without such effect, consensus formation could be difficult. We have considered one such case, which is the question
of the UK leaving the EU or Brexit. This has seen fragmentation of two groups, none of which could achieve 
overwhelming support i.e., consensus was not formed. There is a long history of opinion surveys on this question, which
shows the fluctuating nature of the two opinion groups. Indeed, in this case there could be external noise, bounded confidence 
and even zealots, but a simple model could also give the qualitative agreement of the size distribution of the 
intervals of zero crossing seen in the data and in the model.

In conclusion, we have presented a simple model of kinetic exchange for opinion formation, where two time scales of consensus 
formation are clearly demonstrated. In about 30\% of cases, the model reaches a segregated state, where consensus formation 
takes a much longer time ($\tau_2\sim L^{3.6}$) than the Ising-like coarsening time-scale ($\tau_1\sim L^{2}$). 
 The persistence behaviour and exit probability have been computed to show distinct departure  from the existing results known in two 
dimensions. 
We have compared the fluctuating nature of the order parameter around $0$ in our model having the meta-stable domain-wall states, with the 
opinion survey data for Brexit, also fluctuating between leave and remain majorities, to find a possible agreement in the distribution of the zero-crossing intervals.

Acknowledgements: PS acknowledges financial support from SERB scheme 
EMR/2016/005429 (Government of India).



\begin{thebibliography}{99}
\bibitem{cons} By consensus we mean that the opinions of all the agents are strictly same ($= \pm 1$).
\end{thebibliography}
\end{document}